\documentclass[12 pt]{article}
\usepackage{amsmath,amsbsy,graphicx,amssymb,amsfonts,multirow,multicol,xr,authblk,bm,setspace,xcolor,soul, endfloat}
\usepackage[hyphens]{url} % this needs to happen before the hyperref entry
\usepackage{geometry}
\usepackage{hyperref}
\usepackage{natbib}
\usepackage[runin]{abstract}

\title{Misclassification of Vaccination Status in Electronic Health Records: A Bayesian Approach in Cluster Randomized Trials}
\author[1]{Adam Kaplan}
\author[1]{Collin Calvert}
\author[2]{Bridget C. Griffith}
\author[3,4]{Daniel Bertenthal}
\author[3,4]{Natalie Purcell}
\author[3,4]{Karen Seal}
\author[5,6,7]{Jeffrey M. Pyne}
\author[7]{Karen Anderson Oliver}
\author[8]{Denise Esserman}
\author[1]{David Nelson}
\affil[1]{Center for Care Delivery and Outcomes Research, VA Medical Center, Minneapolis, MN, and Department of Medicine, University of Minnesota, Minneapolis, MN}
\affil[2]{Clinton Health Access Initiative (CHAI), Boston, MA}
\affil[3]{San Francisco Veterans Affairs Health Care System, University of California, San Francisco, San Francisco, CA, USA}
\affil[4]{Departments of Medicine and
 Psychiatry, University of California, San Francisco, San Francisco, CA, USA.}
 \affil[5]{Mental Health Service, Central Arkansas VA Healthcare System, Little Rock, AR, USA.}
 \affil[6]{Department of Psychiatry, University of Arkansas School of Medicine, Little Rock, AR, USA.}
 \affil[7]{Center for Mental Healthcare and Outcomes Research, Central Arkansas Veterans Healthcare System, 2200 Fort Roots Drive, North Little Rock, AR, USA.}
 \affil[8]{Department of Biostatistics, Yale School of Public Health, New Haven, CT, USA.}

\date{\today}

\begin{document}

\maketitle

\section{Introduction}

Misinformation about vaccines and other sociopolitical and individual factors has led to vaccine hesitancy and reduced uptake of immunization in prior epidemics with negative consequences for public health. A recent example is the COVID-19 pandemic in which an unprecedented rapid development and mass roll-out of vaccines heightened public skepticism and concern for safety. Also, newly developed COVID-19 vaccines permeated the national news cycle, as well as social media, and vaccination evolved into a charged political issue. Higher rates of vaccine hesitancy, lack of trust in vaccines, and a reduction in uptake of other adult vaccinations (e.g., influenza) ensued \citep{avarez2021, shet2022impact,zhang2023spillover}. On a national scale, developing novel interventions that address these challenges and increase vaccine uptake rates require access to accurate vaccine record data using electronic health records (EHR) and insurance data. A centralized national or state vaccination registry for all adult vaccination does not exist, and the multiple available sources of EHR data are prone to misclassification, e.g., misclassifying an individual as unvaccinated when having been vaccinated \citep{CDC}.

Misclassification of vaccination status occurs when a study participant’s binary outcome, i.e., vaccinated or not vaccinated, is misclassified. Statistical approaches for measuring the degree of misclassification and ways to address it on an individual level already exist.  Some methods may model an overall misclassification rate or split this rate into false-positive and false-negative rates \citep{xia2018bayesian, liu2017logistic, duffy2004simple}. Others have incorporated steps to allow parameter estimates from observed data to help ‘reassign’ misclassified outcomes. \citet{russo2022robust} flips outliers’ binary values based on a modeled probability of misclassification for all observations and the observed covariates’ relationship with the probability of vaccination. In one outlier example is the model-predicted probability of vaccine uptake for a vaccinated participant is close to 0\% - and in the opposite case the probability is 100\% for an unvaccinated participant. In our experience, the accuracy of this method depends on the base vaccination rate and how strong predictor relationships are with vaccination status. For example, if the base rate is low and predictor relationships are nearly null the correction primarily depends on the prior expected rate of misclassification thereby erroneously suggesting a correction of participants’ statuses.

Misclassifying an individual’s vaccination status can impact the comparison of uptake between a novel behavioral intervention and usual care. Calculating an accurate uptake rate requires identifying the number of individuals eligible for vaccination at the start of the study and the number that received vaccination during the study period. When records are incorrect or inaccessible, some individuals identified as eligible for vaccination are not really eligible (i.e., if they were already vaccinated at the start of the study), and some who did receive a vaccine during the study period would not be identified in the available data (e.g., because the only record of the vaccination is from an unavailable external source). when vaccination status is negatively misclassified in this way, the true vaccination rate is most likely under-estimated by the available data. The ramifications of underestimation can vary. For instance, the novel intervention arm’s observed vaccination rate could be less than that of standard of care but the misclassification rate in the novel study arm is larger. Therefore, accounting for  misclassification is important in gauging the range of estimates for comparative effectiveness between the two interventions.

To accurately measure the impact of an intervention with data collection on a national scale, a pragmatic cluster controlled randomized trial (CCRT) is a promising study design. CCRTs randomize an intervention to groups of people or sites that deploy the intervention to all consented subjects within that group \citep{dron2021role}, in contrast to deploying multiple interventions within one site. The motivation to randomize entire sites rather than individual people within a site is two-fold: to avoid individual randomization (i.e., assigning one person at a time to an intervention arm) when it is difficult (e.g., deploying multiple interventions within the same study site may be arduous); and to avoid the disruption of the clinical practices caused by introducing multiple types of interventions within a single site. A pragmatic CCRT design also allows researchers to tailor intervention implementation to each site that is selected for intervention. Yet routine collection of data across multiple sites increases the potential impact of misclassification. Analysis strategies accounting for binary outcome misclassification in CCRTs are understudied.

Typical analysis of vaccine uptake in a CCRT often fits a binary-outcome regression model to the vaccine status (vaccinated versus not) while accounting for the hierarchical nature of the CCRT via random effects for sites \citep{pinheiro1995approximations, breslow1993approximate, bolker2009generalized}. These generalized linear mixed models (GLMM) account for between-site differences by assuming that each site effect arises from a shared probability distribution. Misclassification still needs to be somehow addressed, and the correction is not readily apparent. The analyst may view misclassification from a missing outcome vantage, in which case they may impute missing binary outcomes within clusters and then fit a GLMM to the imputed data sets and use aggregation methods to summarize parameter estimates \citep{ma2011imputation}. This promising cluster-specific imputation scheme accounts for between-cluster differences but knowing which participants’ vaccination statuses are misclassified (i.e., to designate for imputing) and ignoring estimates of by-facility misclassification rates undervalues the true uncertainty introduced by misclassification. Saliently, this uncertainty can be incorporated via a Bayesian framework by designating the number of participants misclassified within a facility as unknown and can be carefully modeled by incorporating contextual site-level factors (e.g., site size, available summary data on state-level analyses). The proposed method acknowledges the available data may not accurately suggest who was misclassified, but rather will help construct plausible estimates for how many within a facility, were misclassified.

The lack in sensitivity methods for binary outcome misclassification in the context of CCRTs motivates us to propose a novel misclassification addition to Bayesian logistic regression with outcomes fitted on a group level (as in the case of CCRT designs). Section \ref{bayes_logit} introduces Bayesian logistic regression with multiple hierarchical levels (e.g., study participants nested within study sites nested within a healthcare network of sites). Section \ref{deltaprior} introduces  misclassification parameters and means of eliciting their values. Here, the study-team can provide the range of likely values of misclassification rates for each study site, which may be assumed to differ between intervention arms. We perturb the observed counts of vaccines during the study period, and the count of eligible participants, by three different random quantities dictated by the three expert-elicited rates of misclassification at each site. Section \ref{dataapp} describes an application of the proposed method in the context of a behavioral intervention to increase vaccine uptake on a national level among U.S. veterans. The Veterans Health Administration (VHA) is the largest integrated health care system in the United States, providing care at 1,380 health care facilities, including 170 VA Medical Centers and 1,193 outpatient sites of care of varying complexity (VHA outpatient clinics) to over 9.1 million Veterans enrolled in the VA health care program \citep{VHA}. This study was conducted across two VHA regions located in the West and South-Central U.S. Section \ref{sims} describes and discusses numerical experiments evaluating the behavior of this sensitivity method. As a result, we can investigate the impact of the assumed misclassification rates on the estimated odds ratio of vaccination between intervention arms. A discussion in Section \ref{discuss} concludes this work.

\section{Methods}

Our aim is to assess the sensitivity of a vaccination uptake-related intervention effect in a CCRT based on elicited prior assumptions about the rates of misclassification in vaccine status prior to and during the study period. The key assumption is that we know two ways in which the vaccination rates will inaccurately estimated. For site $s$, let $N_{s}$ denote the observed count of eligible participants included for the analysis where inclusion may depend on vaccine status prior to study start, and $Y_{s}$ is the observed count of vaccinations from the eligible pool ($N_{s}$) during the study period. The true vaccination count, $Y_s^*$, and number eligible, $N_s^*$, are inaccurately reported; however, we do have some information that the true vaccination rate, $Y_{s}^{*}/N_{s}^{*}$ is underestimated by $Y_s/N_s$. %Namely, the corrected count of vaccinated participants who were eligible to receiving a vaccination, $Y_s^* = Y_s + \delta_{s,1}$ where $\delta_{s,1}$ is the unaccounted number of vaccinated over the study period. Also, we correct the site size by assuming $N_s^* = N_s - \delta_{s,2} - \delta_{s,3}$, which effectively removes unaccounted ineligible participants split by their observed outcome value, with subscripts equal to 2 and 3 denoting ineligibility among the unvaccinated and vaccinated, respectively. In the simple case, we can assume that ineligibility rates do not differ between the observed vaccination status during the study, implying $\delta_{s,3} = 0$ and redefining $\delta_{s,2}$ to be the number of participants in general that were misclassified as eligible.

%Of course, there are natural bounds on what $\bm{\delta}_s = (\delta_{s,1}, \delta_{s,2}, \delta_{s,3})$. The upper bound for $\delta_{s,1}$ would be the number of eligible unvaccinated. For $\delta_{s,2}$ and $\delta_{s,3}$, upper bounds would be all observed unvaccinated and vaccinated. Without prior \textit{external} knowledge the lower bounds can be set to $0$. If $\bm{\delta}_{s}$ can be ignored, i.e., we fully trust the values of the observed data, then the analysis reduces to a standard comparison of intervention to control sites' vaccinated proportions using GLMM, however the goal is to incorporate our prior beliefs to provide a plausible range of intervention effect estimates arising from various corrections in the observed rates $Y_s / N_s$. 

%In the context of a cluster randomized trial, the intervention effect and the measure of intra-class correlation (ICC) are targets of inference.  In study design, the ICC is used to calculate the sample size given other trial operating characteristics, 
If no misclassification occured then standard analysis using a GLMM with a logistic link function can be fit to the observed data. We review the model below when the observed data is taken as true, i.e., $(Y^*_s = Y_s, N_s^* = N_s, X_{s}, A_{s})$. Then each site's data is modeled as
\begin{equation}
Y^*_s|p_{s} \sim \text{Binomial}(N^*_s, p_s) \hspace{.2cm} \text{with} \hspace{.2cm}
\text{logit}(p_s) = X_{s}^T \alpha + A_{s}^{T}{\theta}_s
\label{eq:GLMR}
\end{equation}
where Binomial$(n,r)$ denotes the Binomial distribution with size $n$ and probability $r$, $\text{logit}(p_s) = \log(p_s) - \log(1-p_s)$, and $p_s$ is the probability of vaccination; $X_{s}$ contains indicators of intervention and possibly site-level covariates that may have been used in stratification during randomization of site to the study arms; $\alpha$ is the vector of effects of these site-level predictors; $A_{s}$ contains the site-specific row of a random effect design matrix $\bm{A}$; and the site-specific random effect vector ${\theta}_s$ are assumed to arise from a multivariate normal distribution commonly centered around 0 with a diagonal covariance matrix. In the case that $\bm{A}$ is the identity matrix we obtain a two-level model with site-level random intercepts, $\theta_{s} \sim N(0, \sigma^{2}_{\theta})$, the latter distribution is the normal distribution centered around 0 with variance $\sigma^{2}_{\theta}$. % For a two-level hierarchy, the ICC is measured by the following estimate $\widehat{ICC}_\theta = \sigma_\theta^2 / (\pi^2/3 + \sigma_\theta^2)$, where the value $\pi^2/3$ is the variance assumed in a logistic distribution, the reference distribution used for logistic regression. In powering a cluster randomized trial it is common to assume that ICC is low in the range of $(0.01, 0.10)$. 

In the current context, the values $Y_s$ and $N_s$ are noisy representations of $Y_s^*$ and $N_s^*$ and a standard GLMM cannot attend to this issue. We review Bayesian logistic regression in Section \ref{bayes_logit} which enables modeling the observed vaccine count and eligible patient count as random quantities. %First, one can assume a fixed value for $\bm{\delta}_s$ for all $S$ sites and proceed with a standard logistic regression; however, this does not reflect the uncertainty in $\bm{\delta}_s$. Instead, one can model $\bm{\delta}_s$ and randomly draw these values for each site. We discuss prior distributions for $\bm{\delta}_{s}$ in Section \ref{deltaprior}.

\subsection{Bayesian Logistic Regression}
\label{bayes_logit}

\citet{polson2013bayesian} introduced the P{\'o}lya--Gamma sampler to facilitate logistic regression in a Bayesian framework. The novelty in this method is that by introducing auxiliary parameters, the initially unwieldy likelihood function for logistic regression can be simplified into two parts: (1) a Gaussian probability model for estimating linear regression coefficients conditional on P{\'o}lya--Gamma variates and (2) another part updating the auxiliary P{\'o}lya--Gamma variates. Coupling this with conjugate prior distributions in the Bayesian framework leads to full conditional posterior distributions to draw samples for the parameters of interest \citep{polson2013bayesian}.

Let $\bm{Z} = [\bm{X} | \bm{A}^{(1)} | \dots | \bm{A}^{(R)}]$ where $\bm{Z}$ is the column-concatenation of the fixed effect matrix $\bm{X}$ and the random effect design matrices $\{\bm{A}^{(r)}\}_{r=1}^{R}$. Next, let \newline $\beta = [\alpha_1,\dots,\alpha_p,\theta_{1,1},\dots,\theta_{1,q_1},\theta_{2,1},\dots,\theta_{2,q_2},\dots,\theta_{R,1},\dots,\theta_{R,q_R}]$, where $p$ is the number of columns in $\bm{X}$, and $q_{r}$ is the number of columns in $\bm{A}^{(r)}$. Per \citet{polson2013bayesian}, we assume \textit{a priori} $\beta \sim \text{MVN}_W(\beta_{0}, \Lambda_{0})$, where $\text{MVN}_{h}(\bm{\mu}, \bm{Q})$ denotes the $h$-variate normal distribution with mean vector $\bm{\mu}$ and covariance matrix $\bm{Q}$, and $W = p + \sum_{r=1}^{R}q_{r}$. Lastly, let $\omega_s$ be the auxiliary P{\'o}lya--Gamma variate for site $s$. With these auxiliary variates, we obtain the following Gibbs sampler steps:
\begin{equation}
\begin{aligned}
    &\omega_s | \beta, N_{s}^{*} \sim \text{P{\'o}lya-Gamma}(N_s^*, Z_s\beta) \\
    &\beta| Y^*_s, N_{s}^{*}, \omega_s \sim \text{MVN}_{W}(\widehat{\beta}, \widehat{\Lambda}).
    \end{aligned}
\end{equation} The $\text{P{\'o}lya-Gamma}$ distribution specified as $\text{P{\'o}lya-Gamma}(a,b)$ has shape and tilt parameters $a$ and $b$, respectively. 
The conditional posterior covariance matrix of $\beta$, $\widehat{\Lambda} = [\bm{Z}^{T}\Omega \bm{Z} + \Lambda_{0}^{-1}]^{-1}$, and mean $\widehat{\beta} = \widehat{\Lambda}(\bm{Z}^{T}\kappa + \Lambda_{0}^{-1}\beta_{0})$. The matrix $\Omega$ is a diagonal matrix with entries $\Omega_{(s,s)} = \omega_s$ and $\kappa = [\kappa_{1},\dots,\kappa_{S}]$ where $\kappa_{s} = Y^{*}_{s} - N_{s}^{*}/2$, for $s=1,\dots,S$. Lastly, we have our data-imputation step by sampling the corrected quantities $(Y^{*}_{s}, N^{*}_{s})$ as laid out in Section \ref{deltaprior}.

Prior distributions are required to fully specify the Bayesian logistic regression with hierarchical components. To align with GLMM methodology, we specify that $\beta_{0} = \bm{0}$ and  $\Lambda_{0}$ is block diagonal with $(R+1)$--many blocks. The first matrix block is a diagonal matrix of size $p\times p$ with diagonal entries $1$ -- this reflects that for the predictors in $\bm{X}$ we have little information on their effect on vaccination uptake but a precision not too large for estimating effects on the probability scale; the remaining $R$--many blocks are diagonal matrices of size $q_{r} \times q_{r}$ each with $\sigma_{r}^{2}$ along their diagonals and $0$ otherwise. Lastly, we specify that each $\sigma_{r}^{-2} \sim \text{Gamma}(0.01, 0.01)$, where $\text{Gamma}(d,e)$ denotes the Gamma distribution with shape and rate parameters $d$ and $e$, respectively. Using this conjugate prior distribution, the full conditional posterior distribution for each $\sigma^{2}_{r}$ is Inverse-Gamma$(0.01 + q_{r}/2, 0.01 + \bm{\theta}_{r}^{T}\bm{\theta}_{r}/2)$, where $\bm{\theta}_{r} = [\theta_{r,1},\dots,\theta_{r,q_{r}}]$ and Inverse-Gamma$(f,g)$ denotes the Inverse-Gamma distribution with shape and scale $f$ and $g$, respectively.   

\subsection{Prior Distribution for Misclassification Rates in Binary Outcomes}
\label{deltaprior}

%In an ideal world we would have information regarding the count of Veterans receiving vaccinations outside and within the VA during the study period. This is not the case. We do know that medicare data has a year lag, and in the context of this project, our available medicare data is one year prior to the study conception. This fact encourages incorporating external data from the past year at the VA while weighing our trust in it to reflect the current year's vaccination rate. While we do not know exactly the number $\delta_{s}$ for all sites $s=1,\dots,S$, we can construct a prior distribution that encapsulates our beliefs about $\delta_{s}$ using past data.

%\hfill \break
%\textit{Truncated Beta Distribution for Unaccounted for Vaccination Rate}
%\hfill \break

Our aim is to correct the observed vaccination counts ($Y_{s}$) and site sizes ($N_{s}$) with corrected counts denoted by $\bm{\delta}_s$ while exploiting the evidence that negative misclassification has occurred, i.e., $N_{s}$ contains more unvaccinated participants than it should and $Y_{s}$ mistakenly contains fewer participants than it should. We define the misclassified counts for site $s$ as $\bm{\delta}_{s} = (\delta_{s,1}, \delta_{s,2}, \delta_{s,3})$; these parameters represent misclassification in the study-period vaccine counts among the eligible $(\delta_{s,1})$, in eligibility of study-period unvaccinated $(\delta_{s,2})$, and in eligibility of study-period vaccinations $(\delta_{s,3})$. For a generic $\delta$, one may proceed by giving a prior distribution among all non-negative integers but the most likely count of misclassified vaccines and related variability about it is difficult to determine for each specific site. Instead, one may assume a rate of misclassification in the unvaccinated during the study reflected by $\rho_{1}$, and in eligibility before the study start denoted by the pair $\{\rho_{2},\rho_{3}\}$, respectively. The misclassification rate in eligibility may differ between the study-attributable unvaccinated (reflected by $\rho_{2}$) and vaccinated ($\rho_{3}$). From this vantage, $(\rho_{2} \times 100)\%$ of the unvaccinated participants were erroneously treated as eligible, and $(\rho_{3} \times 100)\%$ of the vaccinated participants were erroneously treated as eligible. The percent $(\rho_{1}\times 100)\%$ of the eligible participants did attain vaccination status during the study period but were not reported as such according to the available data.

There is a precise order we assume about the misclassification corrections. First, we need to reconstruct the denominator via $N_{s}^{*} = N_{s} - \delta_{s,2} - \delta_{s,3} = N_{s} - \lceil\rho_{2}(N_{s} - Y_{s})\rceil - \lceil\rho_{3}Y_{s}\rceil$, and $E_{s} = Y_{s} -\lceil\rho_{3}Y_{s}\rceil$; this effectively removes ineligible study participants while assuming two different ineligibility rates between study-attributable, observed unvaccinated and vaccinated. Second, $Y_{s}^{*} = E_{s} + \delta_{s,1}= E_{s} + \lceil\rho_{1}(N_{s}^{*} - E_s)\rceil$. The term $\lceil \cdot \rceil$ denotes the ceiling function. This order in corrections implies that $\rho_{1}$ is the rate of misclassification in the study outcome among eligible participants during the study period. To model differential misclassification between intervention arms, we assume that for all $s$ sites in intervention arm $t \in \{1,2\}$ share $\bm{\rho}_{t} = [\rho_{t,1}, \rho_{t,2}, \rho_{t,3}]$. 
\newline

\noindent \textit{Constructing the Prior Distribution for the Misclassification Rate}
\newline

We will refer to a general misclassification rate $\rho$ in the following description of its prior distribution. A natural prior distribution for a random variable that exists from 0 to 1 -- such as $\rho$ -- is the $\text{Beta}(
\gamma, \lambda)$ distribution, where $\text{Beta}(m,n)$ refers to the Beta distribution with shape and scale parameters $m$ and $n$, respectively. Substantial evidence may exist for a lower bound on the misclassification rate to be greater than 0\% and the upper bound is most definitely not 100\% (i.e., not all unvaccinated study participants were misclassified); in any case, lower and upper bounds can be specified \textit{a priori}, let the lower and upper bounds for $\rho$ be $a$ and $b$, respectively. Then we assume the prior distribution for $\rho$, denoted as $\pi(\rho)$, is equal to $\text{Beta}(\gamma, \lambda)_{[a,b]}$, also known as the truncated Beta distribution.

The research team can easily specify $a$ and $b$, but the values of $\gamma$ and $\lambda$ require more information. We can estimate values for these two parameters provided a most likely value for $\rho$ (i.e., the mode), and lower- and upper-end percentiles for $\rho$, denote these as $\kappa_{p_{1}}$ and $\kappa_{p_{2}}$, respectively. Under $\pi(\rho)$, the mode of $\rho$, $M= \frac{\gamma-1}{\gamma+\lambda-2}$; rearranging yields $\lambda = \frac{\gamma(1-M) + (2M -1)}{M}$. Last, using $\kappa_{p_{1}}$ and $\kappa_{p_{2}}$ we can further refine the shape of the desired Beta distribution by minimizing the following objective function: 

\begin{equation}
\Bigg[\int_{\kappa_{p_{2}}}^{b}\text{Beta}(\gamma, \lambda)_{[a,b]}d\rho -
\int_{a}^{\kappa_{p_{1}}}\text{Beta}(\gamma, \lambda)_{[a,b]}d\rho\Bigg] - (p_{2} - p_{1})=0.
\label{eq:objective}
\end{equation}
Specifying the mode changes this estimation procedure from a two-parameter to a one-parameter minimization problem in $\gamma$, and the percentiles aid in constructing the desired shape of a prior distribution for  $\rho$. We encourage $a < \kappa_{p_{1}} < \kappa_{p_{2}} < b$, and $p_{1} < p_{2}$ such that a solution to Equation \ref{eq:objective} can be attained. 

\begin{figure}[h]
    \centering
\includegraphics{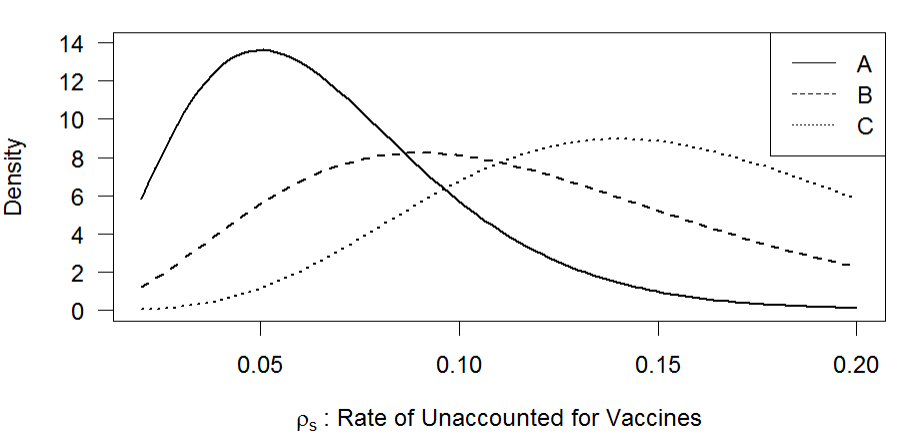}
    \caption[Illustration of Truncated Beta Distribution for Misclassification Rate Prior]{Truncated Beta Distributions to Reflect Prior Belief on Unaccounted for Vaccine Rate. For $a = 0.02$ and $b = 0.20$, the three distributions A, B, and C arise from optimizing Equation \ref{eq:objective} with $(M, \kappa_{p_{1}}, p_{1}, \kappa_{p_{2}}, p_{2})$: A - (0.05, 0.03, 5\%, 0.15, 95\%); B - (0.09, 0.03, 10\%, 0.15, 90\%); C - (0.14, 0.06, 11\%, 0.18, 95\%). }
    \label{fig:truncBeta}
\end{figure}

For example, a research team has two sources of data: the observed data and the observed data with updated vaccination records for some participants but not all. Based on the discrepancies between the two data sources, the team agreed that the most likely value for the unaccounted vaccine rate over the course of their study ($\rho_1$) is around 0.05, the 5th and 95th percentile is determined to be 0.03 and 0.15, respectively, i.e., $(\kappa_{0.05}, \kappa_{0.95}) = (0.03, 0.15)$. The team strongly believes that based on the past year's Medicare vaccination rates, reasonable lower and upper bounds are 0.02 and 0.20, respectively, i.e., $(a, b) = (0.02, 0.20)$. Using the preceding optimization procedure, we obtain distribution A from Figure \ref{fig:truncBeta} for $\rho_{1}$. 

\begin{figure}[h]
    \centering
\includegraphics{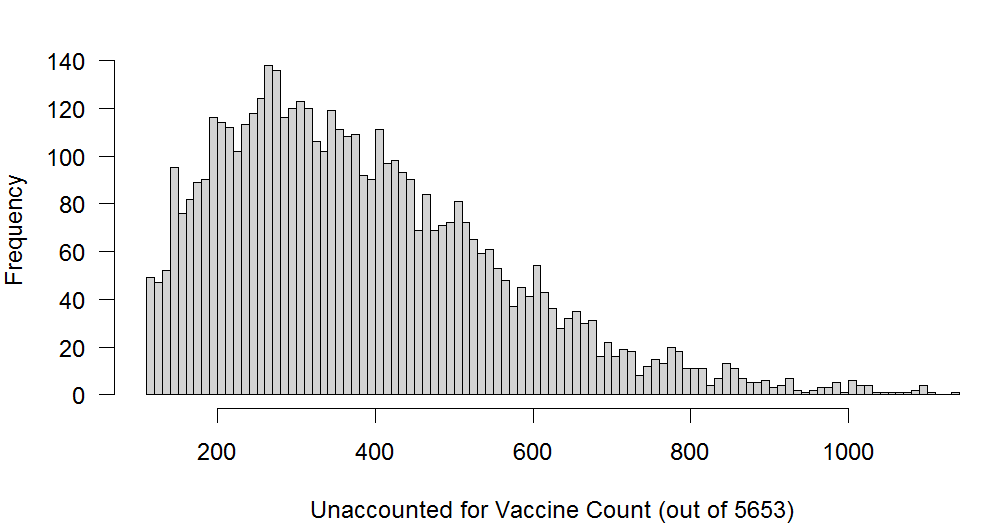}
    \caption[Inferred Prior Distribution for $\delta_{1}$]{Inferred prior distribution for $\delta_{1}$ given the prior distribution $A$ from Figure \ref{fig:truncBeta} and a total number of unvaccinated of $(N - Y) = 5653$. For $a = 0.02$ and $b = 0.20$, distribution $A$ was based on $(M, \kappa_{5\%}, \kappa_{95\%}) = (0.05, 0.03, 0.15)$.}
    \label{fig:truncDelta}
\end{figure}

Distribution A illustrates our assumption that the unaccounted for vaccine rate from a general clinic is most likely 5\% as intended, however, this distribution does allow for decreasing probability for higher values than 5\%. In Figure \ref{fig:truncDelta}, we graph $\delta_{1}$ assuming Distribution $A$ from Figure \ref{fig:truncBeta} based on an observed unvaccinated count of $(N-Y) = 5653$.  

%% write a section on actual data implementation strategies 
\subsection{Gibbs Sampling Steps for Bayesian Logistic Regression with Misclassification Correction Step}

The full conditional Gibbs sampling steps (Section \ref{bayes_logit}) only require the addition of the correction steps for $Y^{*}_{s}$ and $N^{*}_{s}$ (in Section \ref{deltaprior}). We repeat the full conditional steps below and add in the sampling for $\bm{\rho}_{s}$ and evaluate the corrected data values. 

\begin{equation*}
\begin{aligned}
    \omega_s | \beta, N_{s}^{*} &\sim \text{P{\'o}lya-Gamma}(N_s^*, Z_s\beta) \\
    \beta| Y^*_s, N_{s}^{*}, \omega_s &\sim MVN_{W}(\widehat{\beta}, \widehat{\Lambda}),\\
    \sigma^{2}_{r}|q_{r}, \bm{\theta}_{r} &\sim \text{Inverse-Gamma}(0.01 + q_{r}/2, 0.01 + \bm{\theta}_{r}^{T}\bm{\theta}_{r}/2),\\
    \end{aligned}
    \end{equation*}
The expressions for $\widehat{\beta}$ and $\widehat{\Lambda}$ are provided in Section \ref{bayes_logit}. 
\newline 

For each type of misclassification $k = 1,2,3$, and given the site-specific lower and upper bounds, $\{a_{s,k}, b_{s,k}\}$, the lower and upper cumulative percents, $\{p_{(s,k,1)}, p_{(s,k,2)}\}$, and related percentiles, $\{\kappa_{p_{(s,k,1)}}, \kappa_{p_{(s,k,2})}\}$, optimize Equation \ref{eq:objective} with the mode constraint to obtain $(\widehat{\gamma}_{s,k}, \widehat{\lambda}_{s,k})$. Carry this out for each site $s$. For an individual $\rho_{s,k}$, sampling from the prior distribution follows:  

\begin{equation*}
\begin{aligned}
    \rho_{s,k} | \widehat{\gamma}_{s,k}, \widehat{\lambda}_{s,k}, a_{s,k}, b_{s,k} &\sim \text{Beta}(\widehat{\gamma}_{s,k}, \widehat{\lambda}_{s,k})_{[a_{s,k}, b_{s,k}]}.\\
    \end{aligned}
    \end{equation*}
\newline 
Then the data corrections follow this order: 

\begin{enumerate}
  \item $N_{s}^{*} = N_{s} - \lceil\rho_{s,2}(N_{s} - Y_{s})\rceil - \lceil\rho_{s,3}Y_{s}\rceil$ (excluding those erroneously labeled as eligible)
  \item $E_{s} = Y_{s} - \lceil\rho_{s,3}Y_{s}\rceil$ (eligible observed vaccinated count)
  \item $Y_{s}^{*} = E_{s} + \lceil\rho_{s,1}(N^{*}_{s} - E_{s})\rceil$ (reclassifying eligible unvaccinated)
  \item and then reevaluate $\kappa_{s} = Y_{s}^{*}-N_{s}^{*}/2$.
\end{enumerate}

\section{Vaccine Uptake Trial}
\label{dataapp}

The trial that prompted this analysis evaluated the effect of an intervention involving training front-line providers to use a communication technique with their patients, known as motivational interviewing (MI). Using MI, providers can guide individuals toward understanding and expressing their own motivation for a target behavior change, in this case, vaccine acceptance. The study compared sites randomized to have their providers offered MI training for vaccine acceptance versus usual care (UC) (no MI provider training for vaccine acceptance) on vaccination uptake during the one-year follow-up period. Patients were assigned to health care clinics nested within health care systems, warranting a 3-level GLMM. There were 5 health care systems assigned to each intervention arm, with nearly 10 clinics per system. In this CCRT, randomization was at the system level: all providers within a clinic nested in a system randomized to the intervention were encouraged to attend an MI training. The outcome we explore is the receipt of any vaccine dose without eligibility constraints related to vaccinations prior to the study period. %Interventions were randomized on the system level meaning that all providers within a clinic within a system randomized to MI were encouraged to attend MI training, whereas for UC this training was not made available. %MI aimed at improving vaccine-related discourse between provider and participants by aiding the participant to make an informed decision on receiving the vaccine. 
To carry out the proposed misclassification sensitivity analysis, participant-level data was aggregated to clinic-level summaries. For example, instead of a participant's biological sex being used as a covariate adjustment, their associated clinic's proportion of the biological sexes was used. In general, these categorical covariates may have many levels and for each covariate all but one of the level's proportion's should be included in analysis to avoid multicollinearity.

An extant gold standard vaccine record can facilitate assumptions about the rate of misclassification for different clinics. The data associated with the trial is the VHA Administrative data on vaccination records, we refer to these data as VHA data. For this trial, the gold standard was the VHA data supplemented by state level vaccine registries. State level registries were available for 9 out of 10 health care systems. Importantly, coupling the VHA data with the registries' data does not create a perfect vaccination record; however, it provided the study team with lower bounds on negative misclassification rates. Recall that the misclassified individuals are among those labeled in the VHA data as unvaccinated. Project staff agreed that those labeled by the state registries as vaccinated but in the VHA data as unvaccinated comprise the rate of known misclassification among the unvaccinated. We label this proportion as $q_{s,1}$. A second quantity that can provide the upper bound on the plausible misclassification rate is the pool of participants that do not have state registry vaccination records but were reported in the VHA data as unvaccinated, labeled as $q_{s,2}$. We would not expect 100\% of $q_{s,2}$ were misclassified as unvaccinated, and indeed, preliminary evidence suggested that participants without state registry entries were lower utilizers of health care, and in turn, lower utilizers tended to have lower vaccination rates \citep{seal2022association}. Therefore we propose the following equation to set up the sensitivity analysis: 
\begin{equation}
P(\text{To Be Corrected as Vaccinated} | \text{Unvaccinated in Trial Data}) = \widehat{\rho}_{s,1} = q_{s,1} + r \times q_{s,2},
\label{correctionrate}
\end{equation}
where $0 < r << 1$. The value $r$ determines the percentage of the rate of unvaccinated not validated by the state registry we'd assume were in reality vaccinated. The value $\widehat{\rho}_{s,1}$ indicates the specified mode used in optimization of Equation \ref{eq:objective} to elicit a prior distribution for $\rho_{s,1}$.

We carried out the sensitivity analysis with three values of $r$. After determining $r \in \{25\%, 50\%, 66\%\}$, $q_{s,1}$ and $q_{s,2}$ were estimated on a system level due to small numbers on a clinic level. For the one system without state registry records and another system with undesirably low use of state registry, we specified $(q_{s,1}, q_{s,2})$ to be the averages of the other systems' values of these quantities. We labeled the values of $r$ as ``Low"$=25\%$, ``Medium"$=50\%$ and ``High"$=66\%$, and then carried out the misclassification sensitivity analysis by pairing one $r$-label for UC clinics with another $r$-label for MI clinics and calculating their $\widehat{\rho}_{s,1}$ values; this yielded 9 different scenarios (e.g., ``Low" misclassification rates in UC while ``Medium" misclassification rates in MI). Lastly, the four additional parameters eliciting the prior distribution for $\rho_{s,1}$ are needed:  for each site, the required $(\kappa_{0.05}, \kappa_{0.95}) = (\widehat{\rho}_{s,1} - 0.01, \widehat{\rho}_{s,1} + 0.01)$; $(a, b) = (q_{s,1}, q_{s,1} + q_{s,2})$. These specifications imply that most of the plausible values for $\rho_{s,1}$ were tightly within $\pm 1\%$ of $\widehat{\rho}_{s,1}$, with a lower bound set to the empirical misclassification rate based off of the state registries' data. See Table \ref{tab:dataapplication} for the sites' values for $q_{s,1}, q_{s,2}$, ``Low", ``Medium", and ``High" values, and the lower and upper limits used for the sensitivity analysis. 

\begin{table}[h]
    \centering
    \begin{tabular}{ccccccc} 
     &&\multicolumn{2}{c}{External Estimates}&\multicolumn{3}{c}{Mode of $\rho_{s,1}$}\\
   Intervention Arm & System $s$ & $q_{s,1}$ & $q_{s,2}$ & Low & Medium & High \\ 
\hline
   Usual Care & 1 & 0.061 & 0.120 & 0.091 & 0.121 & 0.140 \\
    & 2 & 0.058 & 0.147 & 0.095 & 0.135 & 0.155\\
    & 3 & 0.039 & 0.307 & 0.116 & 0.193 & 0.242  \\
    & 4 & 0.063 & 0.130 & 0.096 & 0.128 & 0.149 \\
    & 5 & 0.031 & 0.113 & 0.059 & 0.088 & 0.106  \\
        \hline
    Motivational Interviewing & 6 & 0.064 & 0.113 & 0.093 & 0.121 & 0.139 \\
    & 7 & 0.064 & 0.092 & 0.087 & 0.110 & 0.125  \\
    & 8 & 0.074 & 0.109 & 0.101 & 0.129 & 0.146  \\
    & 9 & 0.058 & 0.147 & 0.095 & 0.132 & 0.155  \\
    & 10 & 0.063 & 0.115 & 0.092 & 0.121 & 0.139  \\
    \end{tabular}
\caption[Prior Elicitation of Misclassification Rate in Vaccine Uptake Trial]{Prior Elicitation of $\rho_{s,1}$ for Trial Data Application: Given the estimated $q_{s,1}$ and $q_{s,2}$ and a value for $r$ from Equation \ref{correctionrate}, a value for $\widehat{\rho}_{s,1}$ is evaluated in columns ``Low", ``Medium", or ``High". Recall $\widehat{\rho}_{s,1} = q_{s,1} + r\times q_{s,2}$, where $r \in \{0.25, 0.50, 0.66\}$. }
\label{tab:dataapplication}
\end{table}

\begin{figure}
\centering
\includegraphics[scale=.7]{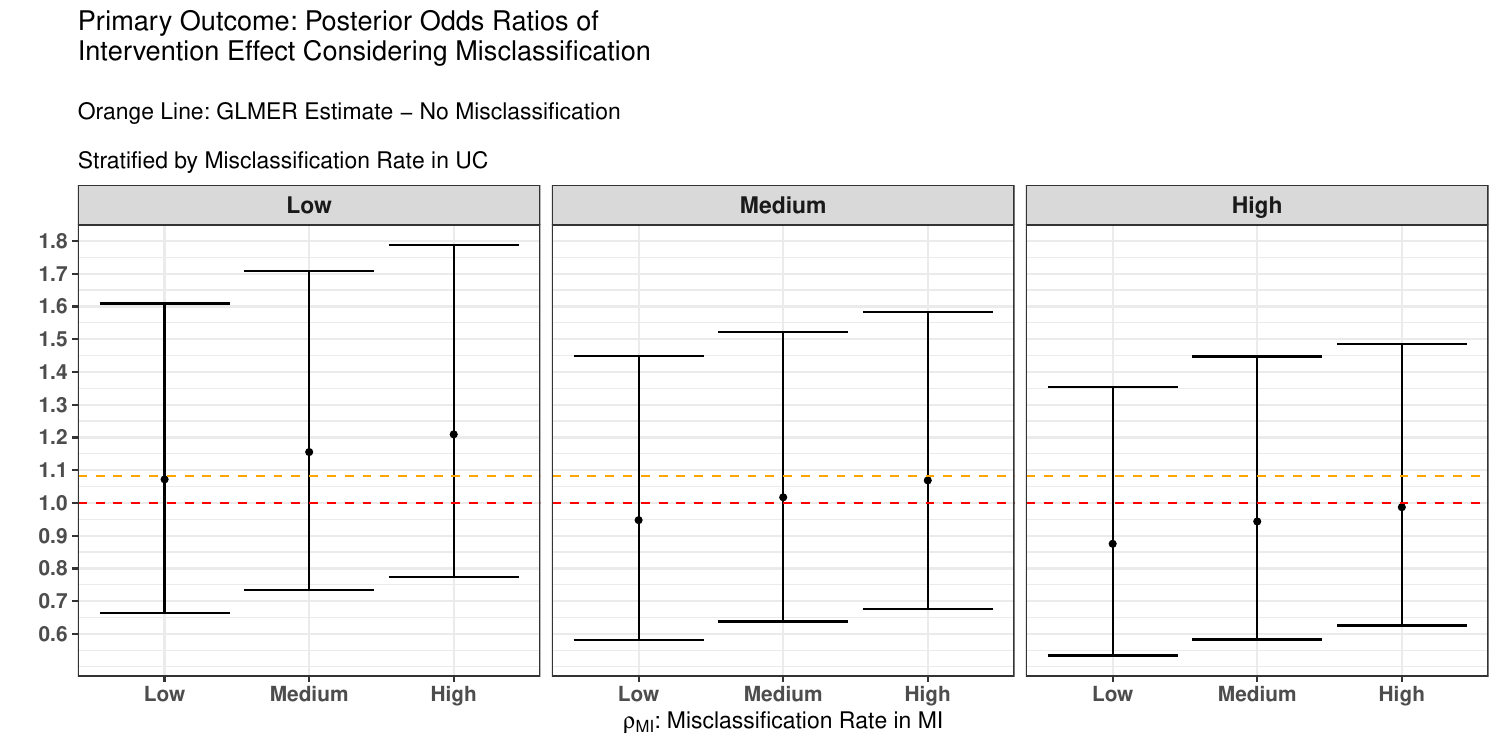}
\caption[Misclassification Correction Sensitivity Analysis Results for Vaccine Uptake Trial]{Vaccine Uptake Trial Application - Varying the degree of assumed misclassification at each health care system differs by intervention arm (see Table \ref{tab:dataapplication}). The odds ratio estimates are odds of any dose in MI over UC. The orange dotted line represents the standard generalized linear mixed model results analyzed on a clinic level for comparison, assuming no misclassification (estimate of 1.082). Numerical values of these estimates are displayed in Table \ref{tab:resultsapplication}.}
\label{fig:applicationresults}

\end{figure}

Figure \ref{fig:applicationresults} displays the results from assuming the degrees of misclassification that differ between the intervention arms listed in Table \ref{tab:dataapplication}. We provide the estimated odds ratio from a standard generalized linear mixed model fit to the observed data and not assuming misclassification corrections (referred to as GLMER). The covariate adjustments were exactly the same between the standard and Bayesian approaches. While the method not assuming misclassification yielded an odds ratio indicating MI had a stronger, yet insignificant, vaccine uptake (estimate of 1.082), the proposed model estimated a wide range of odds ratios across the various misclassification rates. All of the 95\% credible intervals contained the null value of 1.0 indicative of no difference between the arms. We also evaluated the posterior probability that the odds ratio is above 1.0 (shown in Table \ref{tab:biggersims}); in practice, noteworthy odds ratios have posterior probabilities often higher than 95\%. In the most extreme assumption, if the misclassification rates can be assumed to be ``High" in MI systems and ``Low" in UC systems the posterior probability of a non-null odds ratio in favor of MI was 81.24\% which is under the 95\% threshold. 

\begin{table}[h]
    \centering
    \begin{tabular}{cccccc} 
     \multicolumn{2}{c}{Rates of Misclassification $r$} & \multicolumn{2}{c}{Odds Ratio Estimates}&\multicolumn{2}{c}{Prob of $\text{OR} > x$}\\
    \textit{UC} & \textit{MI} & \textit{OR} & \textit{95\% CrI (Lower, Upper)} & \textit{Null (1.0)} & \textit{GLMM OR}  \\ 
 \hline
   Low &Low & 1.072& (0.665, 1.609) & 0.599 & 0.438 \\
  Low & Medium & 1.155 & (0.734, 1.709) & 0.741 & 0.597 \\
  \textbf{Low} & \textbf{High} & \textbf{1.209} & \textbf{(0.773, 1.786)} & \textbf{0.812} & \textbf{0.692} \\
  \hline
  Medium &Low & 0.947& (0.581, 1.449) & 0.353 & 0.222\\
  Medium & Medium & 1.017 & (0.638, 1.523) & 0.490 & 0.334\\
  Medium & High & 1.069 & (0.675, 1.583) & 0.595 & 0.431 \\
  \hline
  \textbf{High} & \textbf{Low} & \textbf{0.875} & \textbf{(0.534, 1.354)} & \textbf{0.223} & \textbf{0.134}\\
  High & Medium & 0.943 & (0.583, 1.447) & 0.338 & 0.214\\
  High & High & 0.987& (0.625, 1.485) & 0.422& 0.277\\
    \end{tabular}
\caption[Misclassification Correction Sensitivity Analysis Results for Vaccine Uptake Trial]{Numerical Estimates of Vaccine Uptake Trial Application of Misclassification Correction Model. Bolded values refer to the assumptions that yielded the highest and lowest odds ratio of vaccine uptake comparing MI to UC, respectively. ``Low", ``Medium", and ``High" refer to the values assumed for misclassification rate modes in Table \ref{tab:dataapplication}. The GLMM estimate refers to an odds ratio of 1.082 from the standard generalized linear mixed model fit to the observed data on a site-level, assuming no misclassification. `Prob of $\text{OR} > x$' refers to the posterior probability of the odds ratio being larger than the specified value of 1.0 (null) or the generalized linear mixed model estimate.}
\label{tab:resultsapplication}
\end{table}

After discussion, the trial team inferred that the ``High" misclassification rates were not representative of the behavior of patients not validated to be unvaccinated: the percentage of these patients assumed to be misclassified would not be as extreme as 66\% because most of these patients seldom interact with VA healthcare systems where exposure to the behavioral intervention would have occurred. In fact, the adjustment for health care system utilization, represented by counts of primary care visits, suggested that lower utilizers associated with decreased chance of vaccination during the study. Therefore, ``Low" to ``Medium" rates of misclassification in the intervention arms were more realistic, implying odds ratios ranging from 0.943 to 1.155 were more reasonable.

\section{Numerical Experiments}
\label{sims}
These sets of simulations evaluate estimation behavior for three scenarios of misclassification with a null to non-null intervention effects based on preliminary data of the COVID-19 vaccine uptake CCRT that motivated this analysis method. We assume that the present trial is a 1:1 randomization scheme with balance in clinic-level covariates post-randomization. There are $S = 90$ study clinics each clinic is assigned to one of $V = 10$ health care systems (HCS). Denote $u_{v}$ as the number of clinics assigned to HCS $v$. The data matrix $\bm{X}$ has a first column pertaining to the intercept term, $X_{s,1} = 1$ for all $S$ clinics. We ordered the next column by control and then intervention sites by specifying that the first half of entries in the second column, $X_{s,2}$ be set to 0 (denoting control clinics), and 1 elsewhere. The vector $\alpha = [\alpha_{1}, \alpha_{2}]$ are the only all-clinic shared effects, i.e., $p=2$; these two effects correspond to $X_{.,1}$ and $X_{.,2}$. For the random effect design matrices we model a nested hierarchy, i.e., $\bm{A}^{(1)} = I_{90}$ and $\bm{A}^{(2)} = \text{BDiag}(\bm{J})$ where $\text{BDiag}(\bm{\lambda})$ denotes a block diagonal matrix with elements $\bm{\lambda}$, $\bm{J} = \{J_{u_{1}},\dots, J_{u_{V}}\}$, and $J_{u_{v}}$ is a $u_{v}$-length vector of 1s. Importantly we order the rows of the design matrices of fixed and random effects to be ordered by HCS labels, thus enabling easier contruction of the random effect design matrices.  The between-clinic variance $\sigma^{2}_{1} = 0.25$ and between-HCS variance $\sigma^2_{2} = 0.07$; these random-effect variances were based on the power calculations of the motivating CCRT.

We used the observed $N_{s}$ from the data collected three months prior to primary analysis of the CCRT and assumed an observed 30\% vaccination rate in the control arm, based on national rates during the study period. The observed vaccination rate in the intervention arm was specified to be 33\%; the observed odds ratio (OR) was 1.149. We simplified the assumptions in the true misclassification process by assuming $\rho_{3,t}^{*} = \rho_{2,t}^{*} = 0.04$ for $t=1,2$, i.e., there was no difference in the misclassification rate among the vaccinated and unvaccinated during the study in terms of eligibility. True rates of misclassification between the intervention arms and the true values of the odds ratios are shown in Table \ref{tab:biggersimstruth}. The next steps refer to the data generation scheme. 

For each site the true sample size was drawn $N^{*}_{s}|N_{s},\rho^{*}_{s,2} \sim \text{Binomial}(N_{s}, 1-\rho_{s,2}^{*})$, where $\rho_{s,2}^{*}$ is the true eligibility misclassification rate. Then we set $\bm{h}_{s} = [p_{s,0}, h_{s,1}, h_{s,2}]$ where $p_{s,0} = \text{logit}^{-1}(\alpha_{0} + \alpha_{1}X_{s,2} + \tau_{s,1} + \tau_{v(s),2})$, $h_{s,1} = \rho^{*}_{s,1}(1-p_{s,0})$ and $h_{s,2} = 1-p_{s,0} - h_{s,1}$, and $\rho^{*}_{s,1}$ is the true study-vaccination misclassification rate, $\tau_{s,1}$ and $\tau_{v(s),2}$ represent clinic and HCS random effects, where $v(s)$ denotes HCS $v$ that clinic $s$ is assigned. We drew $\tau_{s,1} \sim N(0, \sigma^{2}_{1})$ and $\tau_{v(s), 2} \sim N(0, \sigma^{2}_{2})$. Then, we drew $\bm{M}_{s} = [M_{s,1}, M_{s,2}, M_{s,3}] \sim \text{Multinomial}(N^{*}_{s}, \bm{h}_{s})$ and set $Y^{*}_{s} = M_{s,1} + M_{s,2}$, where $\text{Multinomial}(B, \bm{u}_{d})$ denotes the multinomial distribution with size $B$ and $d$-length probability vector of elements $\{u_{g}\}^{d}_{g=1}$. On average, the vaccination rate after corrections was $p_{s,0} + \rho^{*}_{s,1}(1-p_{s,0})$ among the eligible patients. Importantly, an observed odds ratio may appear null due to misclassification but in reality the true odds ratio can be over- or under-estimated. This bias depends on the vaccination and misclassification rates in each arm.

We fit the observed data using several methods for comparison: (1) GLMM using {\tt{glmer}} in the R package {\tt{lme4}} fit to the observed data (henceforth GLMER), (2) GLMER fit to the corrected data, (3) Bayesian logistic regression incorporating $\bm{\rho}_{1}$ and $\bm{\rho}_{2}$ with specifications laid out in Table \ref{tab:biggersims}. We collected 2000 posterior samples of parameters after a 500 sample burn-in. We ran 100 Monte Carlo iterations of these simulations and collected the following on the log-odds scale: average bias in the intervention effect, 95\% uncertainty interval coverage and half-width of the true intervention effect; we also collected average bias in true usual care and intervention rates of vaccination on the probability scale. The specifications in the first three rows of Table \ref{tab:biggersims} assumed both arms have equivalent misclassification rates (referred to as Set 1), whereas in the last three rows these specifications assumed differential misclassification between intervention arms (referred to as Set 2). Our hypotheses for this experiment are that by letting the data corrections be random we are allowing for much more variability and will widen the uncertainty intervals in estimating $\alpha_{2}$, and that bias in $\alpha_{2}$ grows as our assumptions in $\bm{\rho}_1$ and $\bm{\rho}_2$ differ from their true values.    

\begin{table}[h]
    \centering
    \begin{tabular}{cccccc} & 
    \multicolumn{2}{c}{Misclassification Rates}  & \multicolumn{3}{c}{True Values} \\
   Scenario \# & $\rho_{1,1}^{*}$ & $\rho_{2,1}^{*}$ & $r_{1}^{*}$ & $r_{2}^{*}$& $OR^*$ \\
    \hline
 I & 0.07 & 0.07 & 0.377 & 0.349 & 1.129\\
II& 0.13 & & 0.417&  & 1.335 \\
III&  0.17 & & 0.444 & & 1.490\\
    \end{tabular}
\caption[True Misclassification and Outcome Rates and Odds Ratios for Simulation Study]{True Misclassification Rates and Odds Ratios in Simulation Studies. The observed OR is 1.149. True odds ratios of vaccination between intervention and control arms is denoted by $OR^{*}$. The true misclassification rates for the intervention and control arms during the study period are $\rho_{1,1}^{*}$ and $\rho_{2,1}^{*}$, respectively. The true eligibility misclassification rates, $\rho_{1,2}^{*} = \rho_{2,2}^{*} = 0.04$.}
\label{tab:biggersimstruth}
\end{table}

\begin{table}[h]
    \centering
    \begin{tabular}{cc}    
    \multicolumn{2}{c}{Without Differential Misclassification}\\
    \hline
Eligibility Misclassification $\rho_{2}$ & Vaccination Misclassification $\rho_{1}$  \\
     &\\
     A (0.04, 0.03, 0.05) & 1 (0.07, 0.06, 0.08) \\
     B (0.07, 0.06, 0.08) & 2 (0.13, 0.12, 0.14) \\
     C (0.10, 0.09, 0.11) & 3 (0.17, 0.16, 0.18)\\
        &\\
\multicolumn{2}{c}{With Differential Misclassification}\\
    \hline
 Intervention $\rho_{1,1}$ & Control $\rho_{2,1}$ \\
 &\\
 A (0.09, 0.08, 0.10) & 1 (0.04, 0.03, 0.05) \\
B (0.13, 0.12, 0.14) & 2 (0.07, 0.06, 0.08) \\
 C (0.17, 0.16, 0.18) & 3 (0.10, 0.09, 0.11)\\
 &\\
\multicolumn{2}{c}{$\rho_{1,2}=\rho_{2,2}$: (0.04, 0.035, 0.045)}\\
       \hline
    \end{tabular}
\caption[Model Specifications for Numerical Simulations]{Assumption on the Distribution of the Misclassification Rates: The mode, 5th and 95th percentile of the prior distributions for $\bm{\rho}$ are displayed in triplets, $(M, \kappa_{5\%}, \kappa_{95\%})$. In the numerical experiments all combinations of $(A, B, C)$ and $(1, 2, 3)$ in both types of misclassification are fit to the simulated data. Set 1 contains 9 model assumptions reflecting non-differential misclassification rates, so the prior distributions assumed for $\rho_{1}$ and $\rho_{2}$ are assumed equal between the intervention arms. In contrast, Set 2 contains 9 different scenarios of differential misclassification where $\rho_{1,1}$ and $\rho_{2,1}$ specifications are unequal, while $\rho_{1,2}$ and $\rho_{2,2}$ share the exact same prior distribution. The limit values $(a,b)$ for $\rho_{1,1}$ and $\rho_{2,1}$ are $M \pm 0.02$, while for $\rho_{1,2}$ and $\rho_{2,2}$ are $(0.03, 0.05)$.}
\label{tab:biggersims}
\end{table}

%\begin{figure}
%    \centering
%\includegraphics{beta_dist_bigsims.PNG}
%    \caption{Illustration of Table \ref{tab:biggersims} Prior Assumptions on $\rho$ without differential misclassification. Solid lines refer to prior distributions 1 and 2 for the vaccination misclassification rate $\rho_{1}$. Dashed lines refer to prior distributions A and B for the eligibility misclassification rate $\rho_{2}$. Note that the bounds on 1 and 2 are (0.02, 0.20) whereas for A and B are (0.0, 0.10).}
%\label{fig:beta_analysis}
%\end{figure}

\subsection{Simulation Results}

\begin{figure}
\centering
\includegraphics[scale=.7]{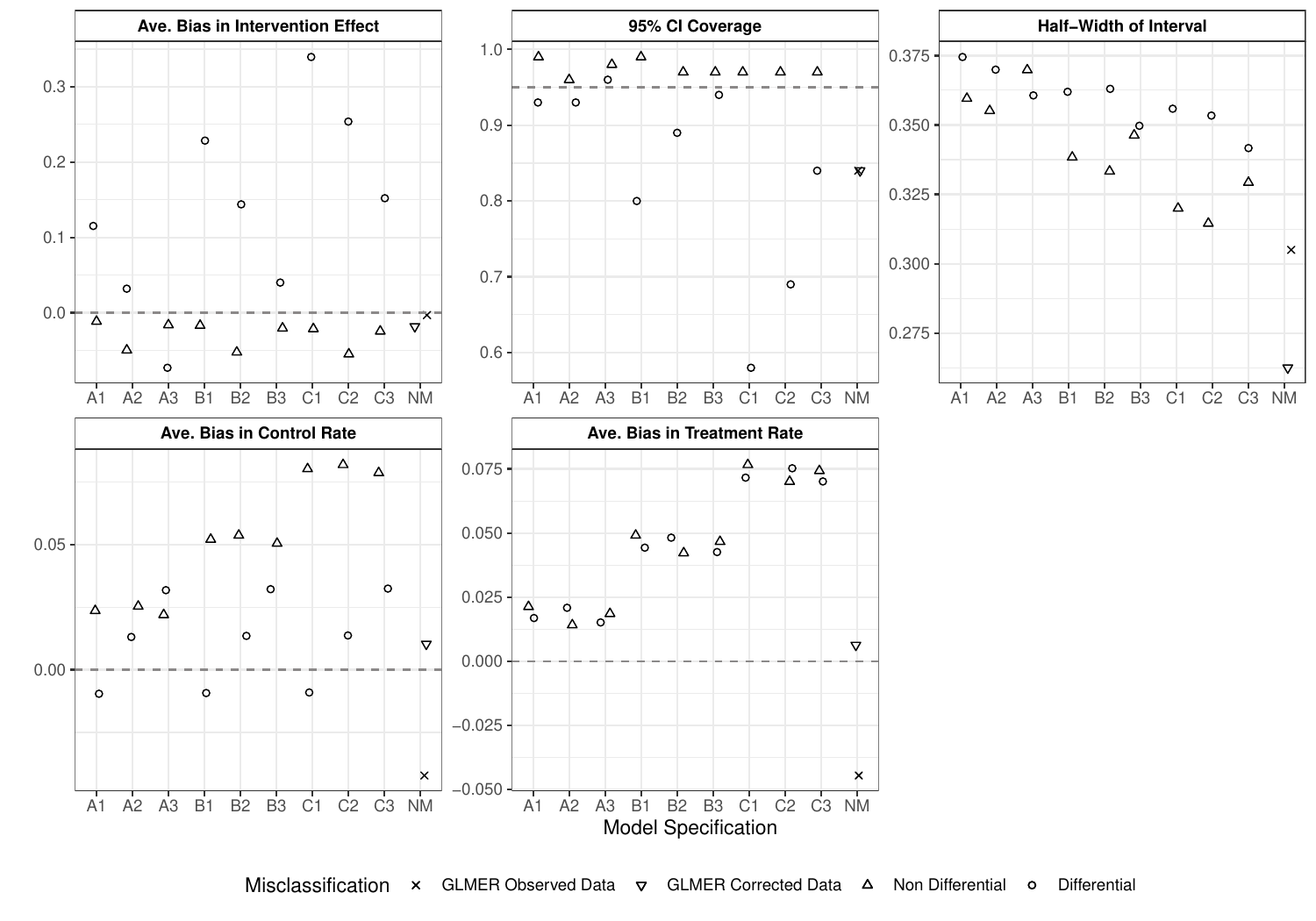}
    \caption[Numerical Simulation Results for Scenario 1]{Simulation Results for Scenario 1 - True odds ratio of 1.129 but overestimated at 1.149 (refer to Table \ref{tab:biggersimstruth}). The label NM refers to Naive Models, GLMER fitted to the observed and corrected data. X-axis labels refer to model specifications laid out in Table \ref{tab:biggersims}.}
\label{fig:SimRes1}
\end{figure}

In Scenario 1, there is no difference in the true misclassification rates between the arms and the true odds ratio is closer to null than the observed. Thus, the specifications that reflect this behavior, most of Set 1, performed similarly well in estimation error (see Figure \ref{fig:SimRes1}). For the same set, increasing the shared $\rho_{2}$ while holding $\rho_{1}$ the same did not alter estimation bias and coverage to a noticeable degree. Set 2 specifications A2 and B3, reflecting similar but not equal rates of outcome misclassification, yielded the lowest bias among Set 2 and close to optimal coverage for both sets; but comparatively lower to Set 1. GLMER with and without the corrected data provided fairly low bias but worse coverage than Set 1. We note that uncertainty interval width decreased as $\rho_{2}$ for Set 1 and as $\rho_{1,1}$ increased for Set 2 specifications; we accredit this behavior to the rates in each arm increasing towards 50\% which associates with the lowest variance of a simple binomial random variable. In turn, as the variability about both rates decrease simultaneously, the variability in the odds ratio decreases as well.

\begin{figure}
\centering
\includegraphics[scale=.7]{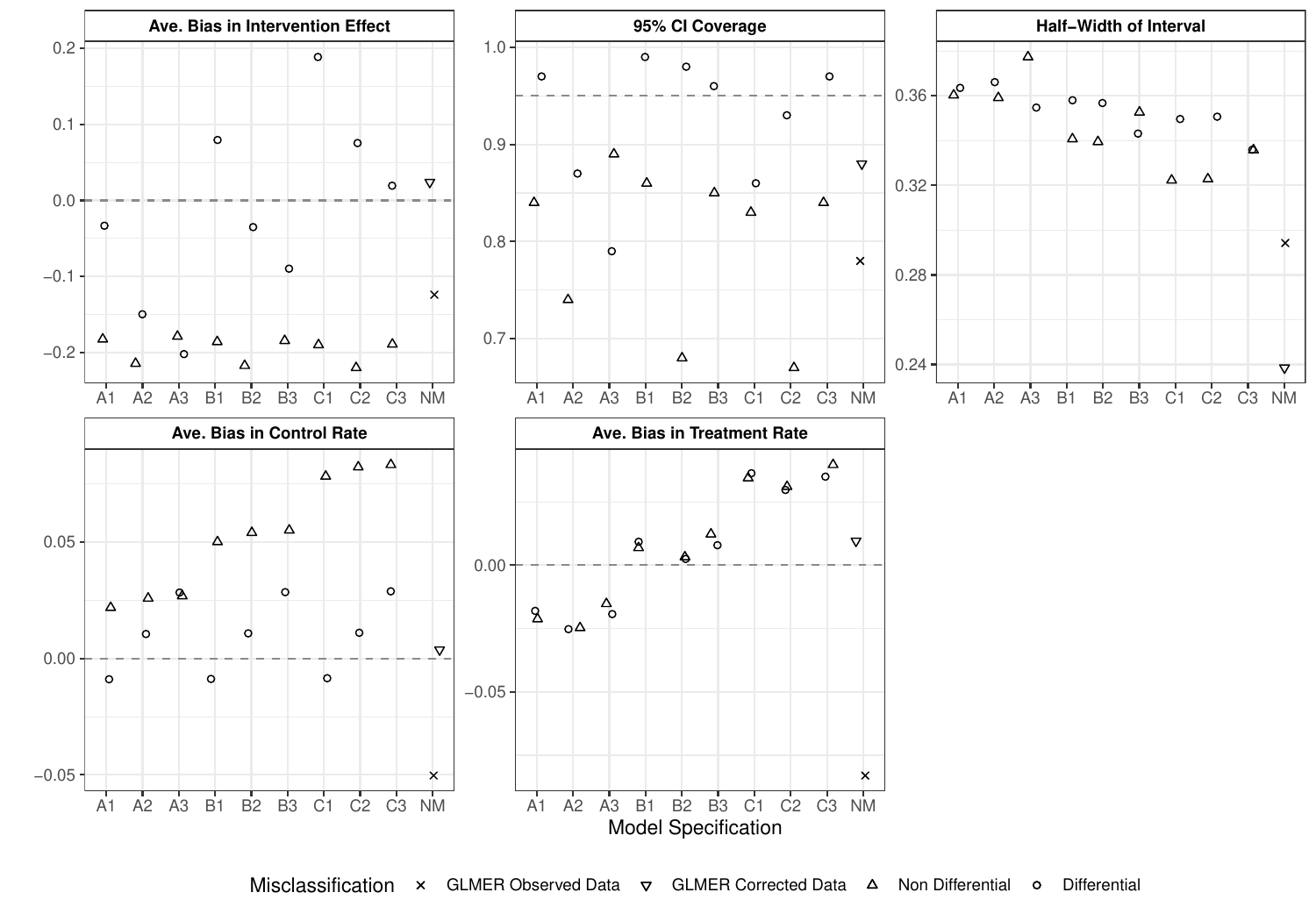}
    \caption[Numerical Simulation Results for Scenario 2]{Simulation Results for Scenario 2 - True odds ratio of 1.335 but underestimated at 1.149 (refer to Table \ref{tab:biggersimstruth}). The label NM refers to Naive Models, GLMER fitted to the observed and corrected data. X-axis labels refer to model specifications laid out in Table \ref{tab:biggersims}.}
\label{fig:SimRes2}
\end{figure}

In Scenario 2, the observed odds ratio is less than the true odds ratio due to the larger misclassification rate in the intervention arm. Set 2 specifications A1, B2, and C3, yielded optimally unbiased odds ratio estimates as compared to all other misclassification approaches (see Figure \ref{fig:SimRes2}). These specifications have differences between $\rho_{1,1}$ and $\rho_{2,1}$ close to the true difference in these rates of 5\%.  Outside of A1, B2, and C3 in Set 2, the other specifications yielded negatively and positively biased estimates. Set 1 specifications yielded negatively biased odds ratio estimates and poor coverage. Not only was interval coverage nearly nominal for Set 2 A1, B2, and C3 but B3 and C2 provided decent coverage of the true effect. In contrast, GLMER modeling the correct data had noticeably lower coverage in comparison.   
%In Scenario 1, the true non-null intervention effect is being overestimated to be close to null using the observed data and the naive approaches agree in capturing this observed odds ratio (Figure \ref{fig:SimRes1_NegativeOR}). In contrast to Naive Bayes, GLMER yielded tighter uncertainty intervals thereby producing worse interval coverage. The non-differential misclassification models yielded positively biased intervention effects, uncertainty coverage near 95\% and the uncertainty interval widths varied across the modeling specifications. Tighter interval widths associated with assuming higher $\rho_{1}$ in both intervention arms, that is, increased rate of vaccination. The differential misclassification approaches yielded a wide spread of biased intervention effects across different modeling assumptions. Model assumption B2 were the closest to the true $\rho_{1}$ values and associated with lower bias and close to optimal credible interval coverage, however, due to the assumptions on $\rho_{2}$ for both arms being equal (counter to the truth), model performance was not optimal. In fact, one could specify C3 assumptions and obtain even better results, while assuming the treatment rate is much larger than it is in truth.

Scenario 3 is similar to Scenario 2 except that the misclassification rate in the intervention arm is much larger than the latter scenario; this yielded a much larger true OR. In Scenario 3, setting C2 from Set 2 provided the lowest estimation error and nearly optimal uncertainty interval coverage. A variety of Set 2 specifications yielded favorable coverage of the true effect. Setting B1, with a difference in $\rho_{2,1}$ and $\rho_{1,1}$ close to the true difference of 9\%, yielded favorable coverage with slightly larger average bias. As expected, the non-differential misclassification methods performed worse in this scenario, quite similar to the estimation behavior of GLMER fit to the observed data. 

\begin{figure}
\centering
\includegraphics[scale=.7]{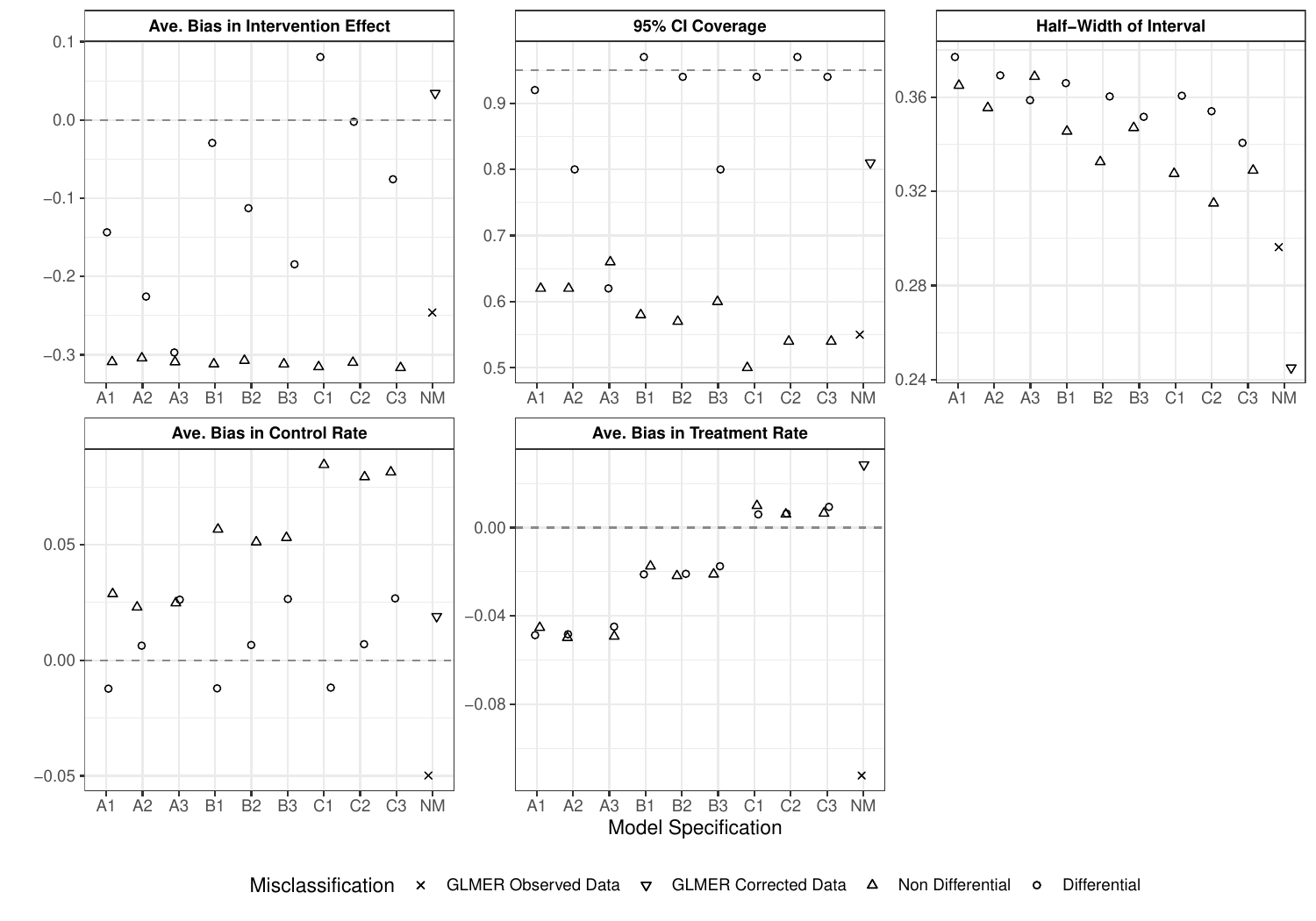}
    \caption[Numerical Simulation Results for Scenario 3]{Simulation Results for Scenario 3 - True odds ratio of 1.490 but underestimated at 1.149 (refer to Table \ref{tab:biggersimstruth}). The label NM refers to Naive Models, GLMER fitted to the observed and corrected data. X-axis labels refer to model specifications laid out in Table \ref{tab:biggersims}.}
\label{fig:SimRes3}
\end{figure}

\section{Discussion}
\label{discuss}

We have introduced a simple noisy-outcome extension of logistic regression and discussed the estimation performance of it in multiple misclassification scenarios. This approach maintains the hierarchical structure of common analysis models for CCRTs while also having strong posterior chain convergence due to the P{\'o}lya-Gamma sampler. We described how to construct prior distributions for group-level units' rates of misclassification given expert-elicited values. Specifying differential rates of misclassification between intervention arms' groups allowed for close recovery of the true intervention effect in our simulation studies when differential misclassification was true. In fact, correct specification of these misclassification rates in each arm was not necessary, but the difference between the specifications being approximately close to their corresponding true difference optimized estimation performance. If differential misclassification is ignorable, then standard GLMM fit without assuming noisy outcomes can be used with little to no loss in parameter estimation, yet the uncertainty interval coverage will be lower as compared to Bayesian approaches.  
 
The proposed method was applied to clinical trial data to investigate that if misclassification rates were different between Motivational Intervewing and Usual Care sites would the effect of MI on vaccine uptake be larger than reported (without assuming differential misclassification). Study investigators may be uncomfortable specifying misclassification rates without thorough research on prior estimates of misclassified vaccination rates. We found a middle ground: we based the lowest value for  misclassification rates on the external, observed rates of misclassification between state-level registries and our collected data for each site. The true unknowns, study sites without vaccination records updated from state-level registries for patients labeled as unvaccinated, provided hypothesized upper limits on negative misclassification rates. This combination of observed and hypothesized rates of misclassification may be useful for different situations outside of vaccination uptake studies: for example, while participants of a particular country in some data sources are promisingly reported via lab tests as having a disease based on latent symptoms, a portion of the study sample may not have access, and thereby not tested at the lab, and otherwise may be misclassified as disease-free. Furthermore, the observable proportions of access to labs may differ between region or city, which correspond directly to the hypothesized proportion of misclassified participants. 

We acknowledge that trading out correcting binary outcomes based on patient-level features (as in \citet{russo2022robust}) with enlisting experts to elicit prior rates of misclassified binary outcomes on a group level may be uncomfortable. An extension of \citet{russo2022robust} for CCRTs could be possible by enabling that the prior rates are group-level specific instead of an overall prior rate of misclassification. However, our epidemiological vantage encourages emphasis on group-level estimation behavior and not invested in learning which participants are incorrectly labeled as vaccinated. The proposed model's computation with simulation data and analysis example can be found at the following website: \url{https://github.com/ankaplan02/Noisy_Binary_CCRT}.

\newpage 
\bibliography{paper2bib.bib}

\clearpage

\noindent{\large{Appendix}}
\newline
\newline

\noindent \textit{Accounting for Ineligibility Misclassification Associates with Reduced Precision in Odds Ratio}
\newline 

We have previously shown that the expected value of the rate of vaccine uptake after misclassification correction in the outcome status for a generic site among the eligible participants is $p_{s,0} + \rho_{1}^{*}(1-p_{s,0})$. It remains to be shown how does $\rho_{2}$, the ineligibility misclassification rate, effects parameter estimation. 

While intuitively known, reducing the sample size for all sites in a CCRT is associated with reduced precision in the intervention effect estimate, we demonstrate that the precision in the odds ratio of vaccine uptake also reduces when accounting for $\rho_{2}$, the ineligibility misclassification. Our key assumption is that ineligibility misclassification does not differentiate between outcome status during the study, i.e., $\rho_{2} = \rho_{3}$. First, we start by recalling the posterior precision of the parameter vector $\beta$, $\widehat{\Lambda} = [\bm{Z}^{T}\Omega \bm{Z} + \Lambda_{0}^{-1}]$. To show that the precision under $\rho_{2}$ is larger than $\rho'_{2}$, for some $\rho'_{2} \neq \rho_{2}$, we show that their difference in precisions is a positive definite matrix for the same data $\bm{Z}$ and coefficients $\beta$. Namely, we need to show that 
\begin{align*}
\widehat{\Lambda}|_{\rho_{2}} - \widehat{\Lambda}|_{\rho'_{2}} &\succcurlyeq \bm{0}, \\
[\bm{Z}^{T}\Omega|_{\rho_{2}} \bm{Z} + \Lambda_{0}^{-1}] - [\bm{Z}^{T}\Omega|_{\rho'_{2}} \bm{Z} + \Lambda_{0}^{-1}] & \succcurlyeq \bm{0}, \\
\bm{Z}^{T}\Big(\Omega|_{\rho_{2}} - \Omega|_{\rho'_{2}}\Big)\bm{Z} & \succcurlyeq \bm{0}. \\
\end{align*}
We can safely assume that the data matrix $\bm{Z}$ is fixed and is non-zero, so we need only worry about the difference between the two $\Omega$ matrices specified at different $\rho_{2}$ values. 

Focusing on the matrix $\Omega$, its entries along the diagonal, $\omega_{s}$ for $s=1,...,S$, follow the updated posterior, $\text{P{\'o}lya-Gamma}(N_s^*, Z_s\beta)$. To investigate how $\rho_{2}$ influences the value of $\omega_{s}$, we will explore the latter's iterated expected value. Following \citet{polson2013bayesian},  

\begin{equation*}
\text{E}[\omega_{s} | \rho_{2}, \beta] = \text{E}_{N_{s}^{*}}[\text{E}(\omega_{s} | N_{s}^{*}, \rho_{2}, \beta)] = \text{E}_{N_{s}^{*}}\Bigg[\frac{N_{s}^{*}\text{tanh}(\frac{1}{2}Z_{s}\beta)}{2Z_{s}\beta}\Bigg],    
\end{equation*}
where tanh($x$) denotes the hyperbolic tangent function. Taking the expectation of $N_{s}^{*}$ yields 
\begin{equation*}
\text{E}[\omega_{s} | \rho_{2}, \beta] = \frac{N_{s}(1-\rho_{2})\text{tanh}(\frac{1}{2}Z_{s}\beta)}{2Z_{s}\beta},  
\end{equation*}
where $N_{s}$ is the observed sample size at site $s$. For one diagonal entry $\Omega$, the difference between the expected precisions specified for two different $\rho_{s}$ is
\begin{align*}
\text{E}[\omega_{s} | \rho_{2}, \beta]  - \text{E}[\omega_{s} | \rho'_{2}, \beta] &= \Bigg(\frac{N_{s}\text{tanh}(\frac{1}{2}Z_{s}\beta)}{2Z_{s}\beta}\Bigg)(1-\rho_{2} - (1-\rho'_{2})) & \\
 &= \Bigg(\frac{N_{s}\text{tanh}(\frac{1}{2}Z_{s}\beta)}{2Z_{s}\beta}\Bigg)(\rho'_{2} - \rho_{2}) > 0.
\end{align*}
We need to ensure that this difference is strictly positive to have, in turn, that $\Omega|_{\rho_{2}} \succcurlyeq  \Omega|_{\rho'_{2}}$. The expression $\text{tanh}(x)/x > 0$ for all $|x| < \infty$, along with $N_{s} > 0$. In fact, $\text{lim}_{x\rightarrow 0}\text{tanh}(x)/x = 1$. We can divide the quantity not concerning $\rho_{2}$ and $\rho'_{2}$ over to the right-hand side to yield $\rho'_{2} > \rho_{2}$. This end result implies the precision in $\beta$, under iterated expectation of $\Omega$, is larger when ineligibility misclassification rates are assumed to be smaller. Importantly, the level of uncertainty in the odds ratio for treatment effect on vaccination increases for assumed higher rates of ineligibility misclassification.

\end{document}